\documentclass[twoside]{article}
\usepackage{amsthm,amssymb,amsmath,quantph,epsfig,setspace}

\newcommand{\gen}[1]{\langle #1 \rangle}   
\newcommand{\abs}[1]{\vert #1 \vert} 
\newcommand{\ket}[1]{\vert #1 \rangle}     
\newcommand{\Integers}{\mathbb{Z}} 
\newtheorem{theorem}{Theorem}
\newtheorem{proposition}[theorem]{Proposition}
\newtheorem{definition}[theorem]{Definition}

\begin{document}

\thispagestyle{empty}

\begin{center} 
{\bf $\:$}\vspace{0mm}

{\LARGE \bf Efficient Quantum Algorithms for the }\vspace{2mm}
{\LARGE \bf Hidden Subgroup Problem over Semi-direct Product Groups}\vspace{5mm}

\centerline{\large
Yoshifumi Inui
\footnote{E-mail: psi@is.s.u-tokyo.ac.jp} \hspace{1mm} and \hspace{1mm} Fran{\c c}ois Le Gall\footnote{E-mail: legall@qci.jst.go.jp}}
\vspace*{0.125truein}
\centerline{\footnotesize\it Department of Computer Science, The University of Tokyo}
\baselineskip=10pt
\centerline{\footnotesize\it 7-3-1 Hongo, Bunkyo-ku, Tokyo 113-0033, Japan}
\vspace*{0.045truein}
\centerline{\footnotesize\it and}
\vspace*{0.045truein}
\centerline{\footnotesize\it ERATO-SORST Quantum Computation and Information Project, JST}
\baselineskip=10pt
\centerline{\footnotesize\it Hongo White Building, 5-28-3 Hongo, Bunkyo-ku, Tokyo 113-0033, Japan   }
\vspace*{0.225truein}

\vspace*{0.21truein}

\setlength{\baselineskip}{11pt}%
      \begin{quotation}
\noindent{\bf Abstract.}\hbox to 0.5\parindent{}
In this paper, we consider the hidden subgroup problem (HSP) over the class of semi-direct product groups   
$\mathbb{Z}_{p^r}\rtimes\mathbb{Z}_q$, for $p$ and $q$ prime. We first present a classification of these 
groups in five classes. Then, we describe a polynomial-time quantum algorithm solving the HSP over all the 
groups of one of these classes: the groups of the form $\mathbb{Z}_{p^r}\rtimes\mathbb{Z}_p$, where $p$ is 
an odd prime. Our algorithm works even in the most general case where the group is presented as a black-box 
group with not necessarily unique encoding. Finally, we extend this result and present an efficient algorithm 
solving the HSP over the groups $\mathbb{Z}^m_{p^r}\rtimes\mathbb{Z}_p$.
\end{quotation}
\end{center} 
\vspace*{10pt}


\section{Introduction and Main Results}   

Almost all the quantum algorithms discovered so far that realize an exponential speed-up with respect to the best known 
classical algorithms can be seen as instances of the Hidden Subgroup Problem (HSP), a problem
that asks to find a subgroup $H$ hidden inside a group $G$. 
In particular the integer factoring problem and the discrete logarithm problem, for which Shor has
presented polynomial-time quantum algorithms \cite{shor}, and
the periodicity finding problem, for which Simon has shown an efficient quantum algorithm \cite{simon},  
are instances of the special case of the HSP where the group $G$ is Abelian. 
More generally, a polyno\-mial-time quantum algorithm solving 
the HSP over any Abelian group $G$ is known \cite{kitaev}, using as its main tool the Fourier transform
over Abelian groups.
However, no solution is known for the general case of $G$ non-Abelian. 
The case of non-Abelian groups is indeed of paramount importance because 
a polynomial-time solution for the HSP when $G$ is the symmetric group (the group of all the permutations over a given set)
would give an efficient quantum algorithm solving the graph isomorphism problem, a well known problem for which no
polynomial-time classical algorithm is known.   
However, the symmetric HSP seems difficult, even for quantum computers, as shown by several negative results 
\cite{hallgren,grigni,moorea,mooreb,hallgren+STOC06}.
Another fundamental instance of the non-Abelian HSP is the case where $G$ is the dihedral group.
Regev \cite{regev} has shown that an efficient algorithm solving the HSP over the dihedral group by the 
coset sampling technique would enable a quantum computer to find, in polynomial time, the shortest vector in a lattice,   
at least for a class of lattices for which no efficient classical algorithm is known. 
Besides the theoretical importance such a quantum algorithm may have, this algorithm would also 
give strong indications that recent cryptosystems proposed by Ajtai and Dwork \cite{ajtai} and Regev \cite{regev2}, 
which are among the best candidates to replace RSA-like cryptosystems and
assume the hardness of computational problems in lattices, are not secure against adversaries using quantum computers.
That is why an important part of the research on the HSP focused on the case where
$G$ is the dihedral group. 
Notice that although no
polynomial-time quantum algorithm is known solving this case,
a quantum algorithm running in sub-exponential time
has been discovered by Kuperberg \cite{kuperberg}, and then improved by Regev \cite{regev3}.     

The dihedral group   
can actually be defined as the semi-direct product $D_{n}=\mathbb{Z}_n\rtimes\mathbb{Z}_2$.   
Ettinger and H{\o}yer~\cite{ettinger} showed that considering the group $D_n$ as the Abelian group
$\mathbb{Z}_n\times \mathbb{Z}_2$, and applying the Abelian Fourier transform over it is sufficient to obtain 
relevant information about the hidden subgroup.
However, the post-processing proposed in \cite{ettinger} requires   
exponential-time to extract a set of generators of this subgroup from this information and thus the global algorithm is not efficient.  
If, for other values of $n$ and $q$,   
the groups $\mathbb{Z}_{n}\rtimes\mathbb{Z}_q$ is sufficiently Abelian, this method or other methods that failed  
to solve completely the dihedral case may work and it is one of the motivation for considering this class of semi-direct product groups.  
Indeed, Moore, Rockmore, Russell and Schulman \cite{moore} proposed a polynomial-time quantum algorithm based on the 
non-Abelian Fourier sampling method solving the HSP over the $q$-hedral groups $\mathbb{Z}_p\rtimes\mathbb{Z}_q$ 
where $p$ and $q$ are two primes such that $q$ divides $p-1$ and $p/q=poly(\log p)$.

Other quantum algorithms are known solving the HSP over some classes of semi-direct product groups  
that are not semi-direct product of cyclic groups.
Using the Fourier transform over $\mathbb{Z}_{p^k}^n\times \mathbb{Z}_2$, Friedl, Ivanyos, Magniez, Santha and Sen \cite{friedl} 
solved in polynomial time the HSP over the groups $\Integers_{p^k}^n\rtimes\Integers_2$ when $p^k$ is a fixed prime power.
Radhakrishnan, R\"otteler and Sen \cite{radhakrishnan} have shown that it is possible to solve in polynomial time, 
information-theoretically, the HSP
over the Heisenberg groups $\Integers_p^2\rtimes\Integers_p$. Another class of semi-direct product groups for which efficient 
quantum algorithms are known corresponds to some wreath product groups \cite{puschel}.

A new promising method has been  recently proposed by
Bacon, Childs and van Dam \cite{bacon}, leading to efficient quantum algorithms solving the HSP over some groups
of the form $A\rtimes\Integers_q$, where $A$ is an Abelian group.
This method is fundamentally different from previous quantum algorithms for the HSP: 
it uses entangled measurements, corresponding to the so-called pretty good measurement,
to identify the hidden subgroup. 
In particular, Bacon, Childs and van Dam's algorithm 
solves in polynomial time the HSP over the groups of the form
$\Integers_{n}\rtimes\Integers_q$, for any integer $n$ and prime $q$ such that $n/q=poly(\log n)$,
thus improving the result \cite{moore}.
They also present an efficient quantum algorithm solving  
the HSP over the $\Integers_p^r\rtimes\Integers_p$, with fixed $r$, improving
the result \cite{radhakrishnan} (and solving completely the problem, not only information-theoretically).

In this paper, we consider the HSP over the class of semi-direct product groups   
$\mathbb{Z}_{p^r}\rtimes\mathbb{Z}_q$, where $p$ and $q$ are prime. The definition of the semi-direct product depending on  
the choice of a homomorphism, we first analyze, in Section \ref{sec:clas}, the different possibilities for this homomorphism in function of   
$p,r$ and $q$. 

Then, in Section \ref{sec:p}, we present a polynomial-time quantum algorithm solving the HSP over the groups of the form
$\mathbb{Z}_{p^r}\rtimes\mathbb{Z}_p$, where $p$ is an odd prime,
even when the group is input as a black-box group with not necessarily unique encoding.
Notice that, prior to our work, the only quantum algorithms for the non-Abelian HSP dealing explicitly with the case 
of black-box groups were the algorithms developed by Ivanyos, Magniez and Santha \cite{Ivanyos+03}. 
In particular, for an arbitrary black-box group it seems usually very difficult to use methods like 
pretty good measurement or Fourier sampling because the explicit form of the generators is unknown.

Although not the
usual setting in HSP research,
studying quantum computation over black-box groups is fundamental for the following reasons.
First, it may be useful in proving separations, in the oracle model (where the oracle is the
black box), of classical and 
quantum computation. 
Second, one of the
most studied case in computational group theory is the setting of permutation groups. 
However, even in this setting, it can happen that factor groups appearing in the computation 
cannot be modeled as permutations groups and can be described only as black-box groups with 
not necessarily unique encoding. Thus, studying the 
HSP in the black-box context (especially with not necessarily unique encoding) may be very useful in order to design 
quantum algorithms for group computational problems over permutation groups as well. 

In Section \ref{sec:gen}, we finally consider the class of groups of the form
$\mathbb{Z}^m_{p^r}\rtimes\mathbb{Z}_p$. Unfortunately, 
the algorithm dealing with the case $m=1$ cannot be generalized easily
and we need other ideas.
We present a quantum algorithm
solving the HSP in polynomial time over these groups 
for any $m$, when the group is input in a special form,
with more restrictions than in the general definition of black-box groups.  

We mention that Chi, Kim and Lee \cite{Chi+06} have recently presented a quantum algorithm, 
based on our results, solving efficiently the HSP over a slightly larger class of semi-direct product groups.
\section{Definitions}
\subsection{The hidden subgroup problem}
We first recall basic definitions and notations we will use in this paper.
For any positive integer $n$, we denote by $\Integers_{n}$ the additive group of integers modulo $n$
and by $\Integers_{n}^\ast$ the multiplicative group consisting of the integers in the set $\{1,\ldots,n-1\}$ that are coprime with $n$.
Given elements $g_1,\ldots,g_s$, we denote by $\gen{g_1,\ldots,g_s}$ the group generated by the generators $g_1,\ldots,g_s$.
Given a group $G$, an element $g\in G$, and a subgroup $H$ we denote
by $gH$ the left coset of $H$, i.e., the set of elements $\{gh\:\vert\:h\in H\}$. 
Now, let us define the notion of an $H$-periodic function.
\begin{definition}
Let $G$ be a group, $H$ a subgroup of $G$ and $X$ a finite set.
A function $f:G\to X$ is said to be $H$-periodic if 
\begin{enumerate}
\item[(i)]
$f$ has the same value on all the   
elements of $G$ in the same (left) coset of $H$, and
\item[(ii)]
$f$ has a different value on each (left) coset of $H$. 
\end{enumerate}
\end{definition}
We now define the hidden subgroup problem.
\begin{definition}
The Hidden Subgroup Problem (HSP) is the following problem. Given as inputs
\begin{itemize}
\item
a group $G$ given as a set of generators, and
\item
a function $f$ given as an oracle,
which is $H$-periodic for an unknown subgroup $H$ of $G$,
\end{itemize} 
output a set of generators for $H$.
\end{definition}
Notice that any group $G$ can be represented by a set of at most $O(\log{\abs{G}})$
generators, where $\abs{G}$ is the number of elements of $G$. We thus say that
an algorithm solves the HSP over $G$ in polynomial time if it runs
in time polynomial in $\log{\abs{G}}$.

\subsection{Semi-direct product groups}
We now define the class of semi-direct product groups of cyclic groups. 

\begin{definition} 
For any positive integers $n$ and $q$, and any group homomorphism $\phi$ from the group $\Integers_q$ into 
the group of automorphisms of $\Integers_n$,    
the semi-direct product group $\mathbb Z_n\rtimes_\phi\mathbb Z_q$ is the set  
$\{(a,b)\:\vert\: a\in \Integers_n,b\in\Integers_q\}$ with the group product  
\begin{displaymath}
(a_1,b_1)(a_2,b_2) := (a_1+\phi(b_1)(a_2),b_1+b_2).
\end{displaymath}
\end{definition} 

Because $\phi$ has to be a homomorphism and $\phi(a)$ must be an automorphism for 
every $a\in\mathbb{Z}_q$, $\phi$ is completely defined by setting $\phi(1)(1)$.   
The group $\mathbb Z_n\rtimes_\phi\mathbb Z_q$ is generated by the two elements $x=(1,0)$ and $y=(0,1)$. 
Using the fact that $\phi(b)(a)=a\phi(1)(1)^b$, we obtain the relation 
\begin{displaymath}
y^{b}x^{a}=x^{a\phi(1)(1)^{b}}y^{b}, 
\end{displaymath} 
which will be used in almost all the group computations in this paper.

\subsection{Black-box groups}

We will mainly consider the case where the group $G$ is input as a black-box group. 
A black-box group is a representation of a group where
elements are represented by strings (of the same length). An oracle that 
performs the group product is available: given two strings representing two elements $a$ and $b$, 
the oracle outputs  the string representing $a\cdot b$. Moreover,
we have another oracle that, given a string representing an element $a$, computes a
string representing the inverse $a^{-1}$.
We will in Section~\ref{sec:p} consider the most general case where  
the elements are not uniquely encoded. 
In this case an  
oracle is provided to check whether two strings represent the same element. 
We refer the reader to Babai and Szemer\'{e}di~\cite{babai} for the complete definition of black-box groups.

In the quantum computation setting, the oracles have to be able to deal with quantum superpositions.
These quantum black-box groups have been studied by Ivanyos, Magniez and Santha \cite{Ivanyos+03} 
and Watrous \cite{WatrousFOCS00,WatrousSTOC01}.
The concept is the same as above but the oracles realizing group operations are
quantum.  More precisely, we suppose that two oracles $V_G$ and $V'_G$ are available, such that
$$V_G(\ket{g}\ket{h})=\ket{g}\ket{gh}$$
$$V'_G(\ket{g}\ket{h})=\ket{g}\ket{g^{-1}h}$$
for any $g$ and $h$ in $G$. 
In the case of a quantum back-box group with not necessarily unique encoding,
we suppose that the oracle checking whether two strings represent the same element 
is a quantum oracle too, although a classical oracle (i.e.~an oracle not dealing with
quantum superpositions) is actually sufficient 
for the algorithms in this paper.

Notice that any efficient black-box algorithm gives rise to an efficient algorithm whenever
the oracle operation can be replaced by efficient procedures. 
Especially, when a mathematical expression of the generators input to the algorithm is known, 
performing group operations can be done directly on the elements in polynomial time (in $\log\abs{G}$) for almost all
natural groups, including permutation groups and matrix groups.
However, we stress that the converse is not
generally true: efficient algorithms for a group problem can use information about the structure 
of the group that are not available in the black-box context.

It is known that the HSP over an Abelian group
input as a (quantum) black-box group with unique encoding can be solved in polynomial time by a quantum computer \cite{mosca2}.
Notice that the same problem is open when the black-box group has not unique encoding.

\section{A Classification of Semi-direct Product Groups}\label{sec:clas} 

\subsection{Number of possibilities for $\phi$}

For given $n$ and $q$, how many possibilities are there for $\phi$ defining a semi-direct product group
$\Integers_{n}\rtimes_\phi\Integers_{q}$?   
The condition that $\phi$ should be a homomorphism implies that $\phi(1)(1)^q\equiv 1\bmod n$.    
Defining $\phi(1)(1)$ satisfying this condition is actually necessary and sufficient to define completely $\phi$.    
Notice that the case $\phi(1)(1)=1$ is a trivial possibility that leads to the direct product    
$\mathbb Z_n\times\mathbb Z_q$.     
By considering the usual decomposition  
\begin{equation}\label{eq:decomposition}
\Integers_n \cong \mathbb Z_{{p_1}^{e_1}}\times\cdots\times\mathbb Z_{{p_k}^{e_k}},
\end{equation}
we can determine the number of possibilities for $\Integers_n\rtimes_{\phi} \Integers_q$ by determining the number of
possible $\phi$ in the definition of the groups
$\Integers_{p_i^{e_i}}\rtimes_{\phi} \Integers_q$. 
Therefore, it is sufficient to study only the case of $n$ being a power of a prime number.  
Finding the number of acceptable definitions for $\phi$ thus reduces to finding elements of order $q$ 
in $\Integers^\ast_{n}$ with $n$ a prime power. In this paper, we will consider only the case $q$ prime,
which gives a clear classification into five classes of semi-direct product groups.

\begin{proposition}\label{propphi}  
Let $p$ and $q$ be two prime numbers, and $r$  an integer such that $r\ge 1$.    
The only cases where there exist non-trivial elements $\alpha$ of order $q$ in $\Integers^\ast_{p^r}$   
are the following three cases.
\begin{itemize}  
\item[(i)] $q\mid p-1$. There are exactly $q-1$ distinct possibilities for $\alpha$.   
\item[(ii)] $r>1,\: q=p\neq 2$. There are exactly $p-1$ distinct possibilities:  
$\alpha=tp^{r-1}+1$ for $0<t<p$.   
\item[(iii)] $r>1,\: q=p=2$. If $r > 2$ then there are exactly three distinct possibilities:  
$2^{r-1}+1$, $2^{r-1}-1$ and $2^r-1$. If $r=2$ then there is only only possibility: $\alpha=3$.  
\end{itemize}
\end{proposition}
\begin{proof}  
First, we consider the case  $p\neq 2$.
Recall that the group $\mathbb Z_{p^r}^\ast$ is a cyclic group.
Let $u$ be a primitive element of $\mathbb Z_{p^r}^\ast$. Then $\alpha$ can be written as $u^k$ for some $k$ less 
than the order of $u$. Since the order of $u$ is $p^{r-1}(p-1)$, $p^{r-1}(p-1)$ divides $kq$.
As $1\le k< p^{r-1}(p-1)$ and we assume $q$ is prime, $q$ must be $p$ or any prime that divides $p-1$.
If $q=p$, $k$ must be of the form $lp^{r-2}(p-1)$ where $l\in\{1,\cdots,p-1\}$, so the number of non-trivial possibilities 
for $\alpha$ is $p-1$. In fact, it can be checked that the order of $\alpha={tp^{r-1}+1}$ is $p$, for every $1\le t\le p-1$: 
these $p-1$ values of $\alpha$ are thus the exact solutions. 
Else if $q$ is a prime that divides $p-1$, $k$ must be $lp^{r-1}\frac{p-1}q$ where $l\in\{1,\cdots,q-1\}$.

Next, we consider the case $p=2$.
Assume $r>2$ (the case $r=2$ is trivial: one unique solution, $\alpha=3$).
As the order of group $\mathbb Z_{2^r}^\ast$ is $2^{r-1}$, the prime $q$ must be 2.
Since $\alpha\in\mathbb Z_{2^r}^\ast$ is odd, we denote $\alpha$ as $2^kl+1$ for $k\in\{1,\cdots,r-1\}$ and odd $l$.
{}From the condition $\alpha^2=2^{k+1}l(2^{k-1}l+1)+1\equiv 1\bmod{2^r}$, we get $k=1$ or $2^r\mid 2^{k+1}$.
We thus obtain three cases:  
the case $k=1$ and $l=2^{r-2}-1$ (corresponding to $\alpha=2^{r-1}-1$), and the case $k=1$ and $l=2^{r-1}-1$ 
(corresponding to $\alpha=2^{r}-1$) and the case $k=r-1$ and $l=1$ (corresponding to $\alpha=2^{r-1}+1$).
\end{proof}

\subsection{Classification of the semi-direct product groups $\Integers_{p^r}\rtimes\Integers_{q}$} 

We have determined the number of possibilities for $\mathbb Z_{p^r}\rtimes_\phi\mathbb Z_q$ as a function 
of $p$ and $q$. 
However, many of these solutions $\phi$ lead to isomorphic semi-direct product groups as stated in the next proposition. 
\begin{proposition}\label{propiso}   
The $q-1$ semi-direct product groups that can be defined using the $q-1$ solutions in the case (i) of Proposition \ref{propphi} are    
isomorphic. Similarly, in the case (ii), the $p-1$ semi-direct product groups corresponding to $\phi(1)(1)=tp^{r-1}+1$    
with $0<t<p$ are isomorphic.   
\end{proposition} 
\begin{proof}  
For the case (i) of Proposition \ref{propphi}, denote by $\phi_1$ one of the homomorphisms. 
The other $q-2$ homomorphisms are actually defined by $\phi_i(1)(1)=\phi_1(1)(1)^i$ for $i\in\{2,\ldots,q-1\}$ coprime 
with $q$. 
We define the one-to-one map $\Psi_i$ from $\mathbb Z_n\rtimes_{\phi_1}\mathbb Z_q$ 
to $\mathbb Z_n\rtimes_{\phi_i}\mathbb Z_q$ by $\Psi_i(x^ay^b):=x^ay^{bi'}$, where $i'$ is the inverse of $i$ 
in $\Integers^\ast_q$. It can be easily checked that  
$\Psi_i(x^ay^bx^{a'}y^{b'})=\Psi_i(x^ay^b)\Psi_i(x^{a'}y^{b'})$. $\Psi_i$ is thus a group 
isomorphism. 

For the case (ii), let $\phi_t$ be the homomorphism corresponding to $\phi_t(1)(1)=tp^{r-1}+1$. 
We define the one-one map $\Psi_t$ from $\mathbb Z_n\rtimes_{\phi_1}\mathbb Z_q$ 
to $\mathbb Z_n\rtimes_{\phi_t}\mathbb Z_q$ by $\Psi_t(x^ay^b):=x^ay^{bt'}$ where $t'$ is the inverse of $t$ 
in $\Integers^\ast_p$. It can be easily checked that  
$\Psi_t(x^ay^bx^{a'}y^{b'})=\Psi_t(x^ay^b)\Psi_t(x^{a'}y^{b'})$. 
\end{proof}   

This implies that there are exactly five classes of non-isomorphic groups $\mathbb Z_{p^r}\rtimes_\phi\mathbb Z_q$, as stated in the
next theorem. 
\begin{theorem}\label{theorem:classification}
The groups of the form $\mathbb Z_{p^r}\rtimes_\phi\mathbb Z_q$, for $p$ and $q$ prime, and $r\ge 1$ can be classified in five non-isomorphic classes:
\vspace{2mm}

\noindent 
{\bf Class 1.}
The q-hedral groups $\mathbb Z_{p^r}\rtimes\mathbb Z_q$ for $p$ and $q$ primes satisfying $q\vert p-1$, and $r\ge 1$;\vspace{2mm} 

\noindent 
{\bf Class 2.} 
The dihedral groups $D_{2^r}=\gen{x,y\:\vert\: x^{2^r}=y^2=e,yx=x^{2^{r}-1}y}$ for $r\ge2$;\vspace{2mm}
 
\noindent 
{\bf Class 3.} 
The quasi-dihedral groups $QD_{2^r}=\gen{x,y\:\vert\: x^{2^r}=y^2=e,yx=x^{2^{r-1}-1}y}$ for $r>2$;\vspace{2mm}

\noindent 
{\bf Class 4.} 
The groups $P_{p,r}=\gen{x,y\:\vert\: x^{p^r}=y^p=e,yx=x^{p^{r-1}+1}y}$ for $p$ prime and $r\ge2$,
except the case $p=r=2$;\vspace{2mm}

\noindent
{\bf Class 5.} 
The direct product groups $\mathbb Z_{p^r}\times\mathbb Z_q$ for $p$ and $q$ prime, and $r\ge 1$.\vspace{2mm}\\ 
\noindent 
Moreover, the five above classes are disjoint.  
\end{theorem} 
\begin{proof} 
Direct consequence of Proposition \ref{propphi} and Proposition \ref{propiso}. 
Class 1 corresponds to the case (i) in Proposition \ref{propphi}.
Class 4 corresponds to the case (ii) and to
the solution $\phi(1)(1)=2^{r-1}+1$ of the case (iii). Class 2 corresponds to $\phi(1)(1)=2^{r}-1$ 
and class 3 to $\phi(1)(1)=2^{r-1}-1$ in the case (iii). Class 5 corresponds to the trivial solution $\phi(1)(1)=1$.
\end{proof} 

\subsection{HSP over semi-direct product groups}

As mentioned above, 
the number of possibilities for the group $\Integers_{n}\rtimes\Integers_{q}$, for $q$ prime, can be obtained 
directly using the decomposition of $\Integers_{n}$ of Equation (\ref{eq:decomposition}).
Notice that the decomposition itself can be found in quantum polynomial time \cite{mosca}. 
However, a subgroup of $G_1\times\cdots\times G_m$ is not necessarily of the form $H_1\times\cdots\times H_m$, 
with $H_i$ subgroup of $G_i$ and, thus, solving the HSP over groups of classes 1 to 5 is not sufficient 
to solve the HSP over any semi-direct product group $\mathbb Z_n\rtimes\mathbb Z_q$.  
But, groups of classes 1 to 5 being basic blocks in the construction of semi-direct product groups, 
we believe it is fundamental
to study the complexity of solving the HSP over groups of each class.

The semi-direct product groups first studied by Moore, Rockmore,
Russell and Schulman \cite{moore} correspond to class 1 with $r=1$. 
These groups are groups of affine functions, 
where the semi-direct product of two elements corresponds to the composition of the associated functions.  
In \cite{moore} a polynomial-time quantum algorithm   
using the so-called strong Fourier sampling method  
was proposed that gives an information-theoretic characterization of any hidden subgroup of this class of group.  
Moreover, when $q$ is sufficiently large, in the sense that $p/q=poly(\log p)$, 
their algorithm returns in polynomial time a set of generators of the hidden subgroup and thus 
completely solves the problem.  

Bacon, Childs and van Dam \cite{bacon} then removed the restriction on $r$ and obtained 
a polynomial time quantum algorithm for the HSP over the groups  $\mathbb Z_{p^r}\rtimes\mathbb Z_q$ (of classes 1 to 5)
when $p^r/q=poly(\log (p^r))$.

The HSP over dihedral groups and quasi-dihedral groups (classes 2 and 3) is one of the most important 
open problem of HSP research.  
In the next section of this paper, we study the semi-direct product groups of class 4 and present a polynomial-time 
quantum algorithm solving the HSP over them.

\section{Quantum Algorithm solving the HSP over $P_{p,r}$} \label{sec:p}  

In this section, we present our quantum algorithm solving, in polynomial time, the HSP over all the groups
of class 4. We recall that, as in Theorem \ref{theorem:classification}, by $P_{p,r}$ we mean the group 
$\gen{x,y\:\vert\: x^{p^r}=y^p=e,yx=x^{p^{r-1}+1}y}$ for $p$ prime and $r\ge2$,
and that the case $p=r=2$ is excluded.

\subsection{Structure of $P_{p,r}$} 
First, using the relation $y^bx^a=x^{a(bp^{r-1}+1)}y^b$, 
it can be easily checked that 
\begin{equation}\label{equation:computation} 
(x^ay^b)^c=x^{a(c+\frac{c(c-1)}{2}bp^{r-1})}y^{bc}
\end{equation} 
for any integers $a,b$ and $c$.
We are now ready to enumerate the different subgroups of $P_{p,r}$.   
\begin{proposition}\label{subgroup}  
The subgroups of $P_{p,r}$ are the following:
\begin{itemize}  
\item
$\langle x^{p^i}\rangle$ for $0\le i\le r$,  
\item
$\langle x^{p^i}, y\rangle$ for $0\le i\le r$,
\item  
$\langle x^{tp^{j}}y\rangle$ with $0\le j<r$ and $1\le t<p$.   
\end{itemize}
\end{proposition}   
\begin{proof} 
For any subgroup $H$ of $P_{p,r}$,     
$H\cap\langle x\rangle$ is of the form $\langle x^{p^i}\rangle$.   
We consider the different possibilities when $H\neq\gen{x^{p^i}}$.  
If $y\in H$ then, necessarily, $H=\gen{{x^{p^i},y}}$.
Suppose otherwise that $y\not\in H$. Then there exists $k\in\{1,\ldots,p^i-1\}$ such that $x^k y\in H$.   
Then
\begin{displaymath} 
(x^ky)^p=\left\{\begin{array}{l} 
x^{kp}\textrm{ if }p\neq 2\\ 
x^{k(2+2^{r-1})}=(x^2)^{(1+2^{r-2})k} \textrm{ if }p=2 
\end{array}\right. 
\end{displaymath} 
and we see that 
\begin{displaymath} 
\gen{(x^ky)^p}=\gen{x^{kp}}, 
\end{displaymath} 
because we do not consider the case $p=r=2$.
This implies that $p^i|kp$ and thus $x^{tp^{i-1}}y\in H$ with $1\le t<p$.
It can be checked that the $p-1$ subgroups $\langle x^{tp^{i-1}}y\rangle$, for $1\le t<p$, are distinct.
\end{proof}   

\begin{proposition}\label{prop_abelian}
All the subgroups of $P_{p,r}$ are Abelian, except the trivial subgroup $\gen{ x^{p^0}, y}=P_{p,r}$.
The only subgroups of $P_{p,r}$ that are not normal are the $p$ subgroups $\gen{ x^{tp^{r-1}} y}$ for 
$0\le t<p$. 
\end{proposition}
\begin{proof}
For the first part, notice that all the subgroups, except the trivial subgroup $P_{p,r}$, contain no element 
of the form $x^ky$ with $1<k<p$ and thus every two elements commute.
We leave to the reader the proof of the second part, tedious but straightforward.
\end{proof}

\subsection{The algorithm}

As shown in Proposition \ref{subgroup}, the group $P_{p,r}$ has $2(r+1)+r(p-1)=O(pr)$ subgroups. 
In the case where $p$ is polynomial in $\log{(p^{r+1})}$, the HSP can be solved 
classically by checking all the subgroups. 
However, this method does not work for general $p$. 

Our algorithm is based on the structure of $P_{p,r}$, resulting from  Proposition \ref{prop_abelian}.
Using this structure,
methods similar to Ettinger-H{\o}yer reduction \cite{ettinger},
can also be used to solve the HSP over $P_{p,r}$, as described by Bacon, Childs and van Dam \cite{bacon}. 
However, our algorithm solves the HSP 
even when the group is input as a black-box group with not necessarily unique encoding. 
More precisely, the problems in this setting are the following. 
With black-box groups, it is difficult, and sometimes impossible, 
to find generators of an arbitrary subgroup and thus Ettinger-H{\o}yer reductions
cannot be directly used. For example, the group $P_{p,r}$ can be input both
by $x$ and $y$ or by $x$ and $xy$. It is thus difficult to find generators 
for a specific subgroup (e.g., the subgroup $\gen{y}$) for an arbitrary black-box representation 
(where the form of the generators is unknown). 
Moreover, when the encoding of the black-box group is not unique, another difficulty appears:
even the quantum Fourier sampling approach to find the order of elements in a group 
cannot be directly used because it is possible that, for example, 
the encodings for the elements of the second period
can be different from the encodings of elements in the first period, although the elements
are the same.

We now present our main result.

\begin{theorem}\label{theo2}
Assume that $P_{p,r}$ is input as a black-box group with not necessarily unique encoding. Then there
exists a quantum algorithm finding, in polynomial time, the hidden subgroup.
\end{theorem}
\begin{proof}
Let $H$ be the subgroup hidden, through the function $f$,
in $P_{p,r}$. 
Any element in $P_{p,r}$ of order $p^r$ is of the form $x^iy^j$ with $p\nmid i$, $0\le  j\le p-1$, 
and any generating set of $P_{p,r}$ contains at least one element of order  
$p^r$ and an element that does not commute with that element. Such two elements can be found 
by testing all the elements of the generating set. 
Let the former be $x^ay^b$ and the latter be $x^{a'}y^{b'}$.
For these elements not to commute with each other, it is necessary and  
sufficient that $ab'\not\equiv a'b\bmod p$. 

There are  
two possibilities, as implied by Proposition \ref{prop_abelian}:\\
\noindent{ Case 1}:
 $H$ is normal in $P_{p,r}$\\
\noindent{ Case 2}:
 $H=\langle x^{tp^{r-1}} y\rangle$ for $0\le t<p$ 

We now present 
two polynomial-time quantum algorithms dealing with each of the 
two cases.
Of course, we do not know which of the 
two cases holds but this does not matter. 
We run the 
two algorithms obtaining
two sets of potential generators for $H$ and output all those that are indeed in $H$ 
(this can be tested by checking whether the value of $f$ on them is $f(e)$).\vspace{2mm}

\noindent{\bf Case 1}:
 $H$ is normal in $P_{p,r}$. \\
We run the algorithm for the normal HSP given by Ivanyos, Magniez and Santha \cite{Ivanyos+03} and
output a set of generators of $H$.\vspace{2mm}

\noindent{\bf Case 2}:
 $H=\langle x^{tp^{r-1}} y\rangle$ for $0\le t<p$ and $b'\not\equiv 0\bmod p$.\\
In this case, $H$ is a subgroup of the Abelian group $\gen{x^{p^{r-1}},y}$. The problem
can be solved easily if explicit generators are known. However, in the case of $P_{p,r}$ being a
black-box group, this is not immediate. 
We show how to find good generators of this subgroup that enable to use the Abelian Fourier sampling method.
Denote $X=x^ay^b$ and $Y=x^{a'}y^{b'}$.
We first find an integer $l$ such that $(X^p)^l=Y^p$. Expanding this using Equation (\ref{equation:computation}) gives
\begin{displaymath} 
\left\{\begin{array}{l} 
al\equiv  a'\bmod{p^{r-1}}\:\:\textrm{    if }p\neq 2\\ 
al(b2^{r-2}+1)\equiv  
a'(b'2^{r-2}+1)\bmod{2^{r-1}} \:\:\textrm{    if }p=2 
\end{array}\right. 
\end{displaymath} 
and guarantees the existence and unicity modulo $p^{r-1}$ of such an $l$.
Now, defining $G'=\Integers_{p^{r-1}}\times\Integers_{p^{r-1}}$, 
$H'=\gen{(l,-1)}$ and $f'(u,v)=f((X^p)^u (Y^p)^v)$, we see that
$f'$ is $H'$--periodic over $G'$. Running the Abelian HSP algorithm enables to
find $H'$ and thus $l$ in polynomial time.

Now, let us first consider the case $p\neq 2$.
By using $l$, we can obtain an element $Y'$ of the form $x^{\alpha p^{r-1}}y^\beta$ where $\beta\not\equiv 0\bmod p$: 
$$Y'=X^{-l}Y=x^{\alpha p^{r-1}}y^{b'-bl}$$ where  
\begin{displaymath}
\alpha p^{r-1}=a\frac{-l(-l-1)}{2}bp^{r-1}-al+a'-a'blp^{r-1}.
\end{displaymath} 
Notice that $b'-bl$ cannot be a multiple of $p$ because, since $a'\equiv al\bmod p$, this would contradict 
the hypothesis $ab'\not\equiv a'b\bmod p$. 
Thus,
$$\gen{X^{p^{r-1}},Y'}= \gen{x^{p^{r-1}},y}\cong \Integers_p\times\Integers_p$$ 
is an Abelian group (see Proposition \ref{prop_abelian}) and
$\langle x^{tp^{r-1}} y\rangle$ is a subgroup of it. We thus use  Abelian
Fourier sampling over $\gen{X^{p^{r-1}},Y'}$ and output a set of generators for $H$.

If $p=2$, then, by a similar argument, 
$Y'=X^{-l}Y$ is of the form $x^{\alpha p^{r-2}}y^\beta$ where $\beta\not\equiv 0\bmod p$. Then
$$\gen{X^{p^{r-2}},Y'}= \gen{x^{p^{r-2}},y}\cong\Integers_{p^2}\times\Integers_p$$ 
is an Abelian group (here, we use our convention that the case $p=r=2$ is excluded) and
$\langle x^{tp^{r-1}} y\rangle$ is a subgroup of it. 
We thus use Abelian
Fourier sampling over $\gen{X^{p^{r-2}},Y'}$ and output a set of generators of $H$.
\end{proof}

\section{Algorithm Solving the HSP over $\mathbb{Z}^m_{p^r}\rtimes\mathbb{Z}_p$}\label{sec:gen}

We finally present a quantum algorithm solving the HSP algorithm over
the groups of the form $\mathbb Z_{p^r}^m\rtimes\mathbb Z_p$, for $p$ prime, where $\Integers_p$ acts separately on each
coordinate of $\Integers_{p^r}^m$ as in $P_{p,r}$. 
Formally, $\mathbb Z_{p^r}^m\rtimes\mathbb Z_p$
is the group generated by
$m+1$ elements $x_1,\ldots,x_m$ and $y$, where $\gen{x_1,\ldots,x_m}\cong\Integers_{p^r}^m$ and
$yx_i=x_i^{p^{r-1}+1}y$ for each $i\in\{1,\ldots,m\}$.

We will now show that, although $\mathbb Z_{p^r}^m\rtimes\mathbb Z_p$ is not an Abelian 
group, applying the Abelian Fourier transform to it
(i.e., the Fourier transform over to the group $\mathbb Z_{p^r}^m\times\mathbb Z_p$) 
is sufficient to get enough information to find the hidden subgroup.

We first state a general useful proposition.

\begin{proposition}\label{proposition:general}
Let $G$ be a black-box group, $H$ a hidden subgroup of G and
$f$ an $H$-periodic function. If there exists a group $G'$ over which 
a quantum polynomial-time solution for the HSP is known 
and a bijection $\pi:G\to G'$ verifying the following conditions
\begin{enumerate}
\item[(i)]
$\pi(H)$ is a subgroup of $G'$;
\item[(ii)]
$f\circ \pi^{-1}$ is $\pi(H)$-periodic;
\item[(iii)]
there is a polynomial-size quantum circuit that, for any $g'\in G'$, maps $\ket{g'}$
to $\ket{w_{g'}}$, where $w_{g'}$ is a string representing
$\pi^{-1}(g')$ (in the black-box representation of $G$); 
\end{enumerate}
then $H$ can be found by a quantum computer in polynomial-time.
\end{proposition}
\begin{proof}
The algorithm for the HSP over $G'$ is used with, as input, the $\pi(H)$-periodic function
$f\circ \pi^{-1}$. This gives, in polynomial time, a set of generators for $H'$.
This set is used to create random elements of $H'$ using standard methods \cite{babai2,babai3},
which are then mapped using $\pi^{-1}$ to obtain almost uniformly random elements of $H$.
A polynomial number of such elements is, with high probability, a
generating set for $H$.
\end{proof}

The above proposition is stated in the most general context of $G$ being a black-box group but,
in this case, the condition (iii) is problematic.
Indeed, even if such a bijection $\pi^{-1}$
exists, it seems very difficult to implement it when the explicit form of the generators for $G$ are
unknown. For $\mathbb Z_{p^r}^m\rtimes\mathbb Z_p$,
we do not know how to do this in the black-box framework and we need to have some 
knowledge of the form of the generators. 
We will solve the HSP over these groups when the input is
given as a set of generators for $\Integers_{p^r}^m$ and one generator for $\Integers_p$. This means that
we can isolate the left part and the right part of the semi-direct product. More precisely, our result is as follows.

\begin{theorem}
Consider $\Integers_{p^r}^m\rtimes\Integers_p$ being input as a black-box group with unique encoding 
under the form
of a set of generators of $\Integers_{p^r}^m\rtimes\{0\} $ and a generator of $\{0\}\rtimes\Integers_p$.
Then there is a polynomial-time quantum algorithm finding the hidden subgroup $H$.
\end{theorem}
\begin{proof}

Denote $A=\Integers_{p^r}^m\rtimes\{0\}$, $y$ the generator of $\Integers_m$ and
$G'=\Integers_{p^r}^m\times\Integers_p=\gen{z_1,\ldots,z_m,z_{m+1}}$.
{}From the set of generator of $A$, we compute a minimal set of generators $\gen{g_1,\ldots,g_m}$ of 
$A$, i.e., $m$ generators that generate $A$. Notice that this is possible, using the algorithm for 
Abelian membership testing by Ivanyos, Magniez and Santha \cite{Ivanyos+03}, because the encoding is unique. 
Let $\pi$ be the following one-one map between $\Integers_{p^r}^m\rtimes\Integers_p$ and $G'$.
$$\pi:g_1^{a_1}\cdots g_m^{a_m}y^b \longmapsto z_1^{a_1}\cdots z_m^{a_m}z_{m+1}^b.$$
This map satisfies condition (iii) of Proposition \ref{proposition:general} because $g_1,\ldots,g_m,y$ are known.

We now prove that (i) holds too.
First, notice that, for any subgroup $H$ of $\Integers_{p^r}^m\rtimes\Integers_p$,
there are two possibilities: $H$ is a subgroup of $ A$ or
$H=\gen{H\cap A, gy}$ for some $g\in A$. 
Indeed, suppose that $H$ cannot be written under the form $H=\gen{H\cap A, gy}$.
This implies that $H=\gen{H\cap A,g_1y,\ldots,g_k y}$ for  $g_1,\ldots, g_k\in A$, with $k>1$.
Then, the elements $(g_iy)^{-1}=y^{p-1}g_i^{-1}$ are in $H$ too.
Thus $g_1g_i^{-1}\in H\cap A$ for all $i\in\{2,\ldots,k\}$, and $\gen{H\cap A, g_1y,\ldots,g_ky}=\gen{H\cap A, g_1y}$, 
which leads to a contradiction. 

If $H$ is a subgroup of $A$, then (i) holds trivially.
Consider the case $H=\gen{H\cap A, gy}$.
{}From Equation (\ref{equation:computation}), for any integer $c$,
$$(gy)^c=g^{\frac{c(c-1)}{2}p^{r-1}}g^c y^c.$$
{}From the fact $g^p\in \gen{(gy)^p}\le H\cap A$, we obtain that $g^{\frac{c(c-1)}{2}p^{r-1}}\in H\cap A$.
Thus $H$ is the subgroup constituted by all the elements of the form $g'g^cy^c$ where $g'\in H\cap A$ and 
$c\in\{0,\cdots,p-1\}$, and $$\pi(H)=\gen{\pi(H\cap A),\pi(gy)},$$ which is a subgroup of $G'$. 
This proves condition (i).

A similar argument proves that
any coset of $H$ in $\Integers_{p^r}^m\rtimes\Integers_p$ is mapped into a coset of $\pi(H)$ in 
$G'$, and more
precisely any two identical cosets are mapped into
identical cosets.  Thus, $f\circ \pi^{-1}$ is $\pi(H)$--periodic and (ii) holds as well.

The HSP over the Abelian group $G'$ can be solved in polynomial time by a quantum computer. 
Using Proposition \ref{proposition:general}, we obtain a polynomial-time quantum algorithm
solving the hidden subgroup problem over $\Integers_{p^r}^m\rtimes\Integers_p$.
\end{proof}


\end{document}